\documentclass[11pt]{article}
\usepackage[normalem]{ulem}
\pdfoutput=1 

\usepackage{jheppub} 
\usepackage{url}
\usepackage{caption}
\usepackage{subcaption}
\usepackage{extarrows}

\usepackage[latin9]{inputenc}
\usepackage{float}
\usepackage{amsmath}
\usepackage{amssymb}
\usepackage{graphicx}
\usepackage{esint}
\usepackage{hyperref}
\usepackage{comment}
\usepackage{color}
\usepackage{microtype}
\usepackage{cleveref}
\usepackage{breakurl}
\usepackage{bbm}
\newcommand{\be}{\begin{equation}}
\newcommand{\ee}{\end{equation}}
\newcommand{\ben}{\begin{displaymath}}
\newcommand{\een}{\end{displaymath}}
\newcommand{\bea}{\begin{eqnarray}}
\newcommand{\eea}{\end{eqnarray}}

   \newcommand{\rf}[1]{(\ref{#1})}
\newcommand{\vp}{\varphi}

\def\be{\begin{equation}}
\def\ee{\end{equation}}
\def\bea{\begin{eqnarray}}
\def\eea{\end{eqnarray}}
\def\ba{\begin{array}}
\def\ea{\end{array}}
\def\bit{\begin{itemize}}
\def\eit{\end{itemize}}

\def\a{\alpha}

\def\vp{\varphi}

 \makeatletter

\allowdisplaybreaks

\makeatother

\makeatletter
\DeclareRobustCommand{\rcite}[1]{%
  \rcite@aux#1,\@nil{#1}%
}
\def\rcite@aux#1,#2\@nil#3{%
  \if\relax#2\relax
    Ref.~\cite{#3}%
  \else
    Refs.~\cite{#3}%
  \fi
}
\makeatother

\hypersetup{
    colorlinks = true,
    citecolor = {blue},
    linkcolor = {blue},
    urlcolor = {blue},
}

 \title{\rm { \LARGE \bf  Alexei Starobinsky and Modern Cosmology
}}

\author{Andrei Linde}

\affiliation{Leinweber Institute for Theoretical Physics at Stanford, 382 Via Pueblo, Stanford, CA 94305, USA}

\emailAdd{alinde@stanford.edu}

\abstract{Alexei Starobinsky is one of the main authors of inflationary cosmology.  Here I will discuss the Starobinsky model and its generalizations, including the theory of $\alpha$-attractors. I will then describe the current status of these models in light of the latest observational results from ACT, SPT, and DESI.}

\begin{document}

\maketitle

 
\parskip 6.5pt

\section{Introduction}

It's hard for me to believe that time runs so fast, and that Alexei (Alesha) Starobinsky and I met each other almost 60  years ago. In 1966,  we became students at the Department of Physics of Moscow State University. For a while, our interests did not intersect. I was interested in quantum field theory and particle physics, working with Kirzhnits.  Starobinsky had a wonderful start investigating particle production in external gravitational fields, in collaboration with Zeldovich. I recall telling my parents at the time that I was very interested in studying cosmology, but ``I'm too old to pursue it.''

And then things changed. Since the 1970s, I have been studying cosmological phase transitions, false vacuum decay, and the transfer of vacuum energy to matter energy, which later formed the basis of both old and new inflation. Meanwhile, Starobinsky studied quantum
effects near the cosmological singularity and at the end of the 70's, he published two prolific papers. In the first one, he described the theory of creation of inflationary gravitational waves  \cite{Starobinsky:1979ty}. In the second paper, he developed the famous Starobinsky model of cosmic inflation, "A New Type of Isotropic Cosmological Models Without Singularity" \cite{Starobinsky:1980te}. 

What makes these papers especially interesting is that they were written before the inflationary theory was proposed by Alan Guth \cite{Guth:1980zm}, and well before the development of the new inflation \cite{Linde:1981mu} and chaotic inflation \cite{Linde:1983gd}. 

The Starobinsky model \cite{Starobinsky:1980te} was an instant sensation among Russian cosmologists back in 1980, but outside of Russia, it was soon overshadowed by many other inflationary models.  Admittedly, he did not attempt to provide a solution to the homogeneity, isotropy, horizon, and flatness problems in these papers, which was the trademark of inflation. But the results of the theory of inflationary perturbations developed by Mukhanov and Chibisov \cite{Mukhanov:1981xt}  in the context of the Starobinsky model provided a perfect match to the WMAP and Planck data \cite{Planck:2018jri}. This model has required only some relatively minor modifications since its introduction 45 years ago, and it still remains one of the most popular inflationary models. 

At that time, I was fully focused on the new inflationary scenario \cite{Linde:1981mu}. I invented it in early Summer 1981.  I spent a few months trying to obtain permission for its publication, and gave a talk about it at the international conference in Moscow in October. The next day after my talk, I was asked to translate a talk by Stephen Hawking at the Sternberg Astronomical Institute. In his talk, Stephen was explaining why the scenario proposed by Guth does not work. And then he said that Linde recently proposed an improved version of this scenario, but it does not work either. For half an hour, I was translating into Russian a detailed explanation of why inflation cannot be improved. After that, I said that I translated, but I disagree. We had a great discussion, and he invited me to the Nuffield Workshop on inflation in the Summer of 1982 in Cambridge.

Preparing for the conference, I issued four new papers on new inflation. About a month before the Cambridge conference, another conference on cosmology took place in Tartu, Estonia. I gave a talk about new inflation there, and then... Starobinsky gave a talk claiming that inflationary perturbations in new inflation were too large, which effectively ruled out this theory.

Well, first, Hawking claimed that new inflation is dead, then Starobinsky also claimed that it is dead, but for a different reason. With friends like that... However, soon after I returned from Tartu to Moscow, I received a preprint by Hawking, which claimed that the amplitude of inflationary perturbations in new inflation is fine. That was getting really interesting.  

The three-week-long Nuffield Workshop was the most significant conference of my life. This was the time when the status of inflationary cosmology was fully recognized. This included the status of the theory of inflationary perturbations in the context of the new inflation scenario. I will explain it by the words of Alesha Starobinsky from his message to me from 2021: 

``{\it I firmly remember that just after my arrival in
Cambridge, I saw the first version of Steven's 1982 paper, and when I had a short meeting with him soon (probably just on the next day after
arrival), I told him that I had got a different (and large for the Higgs-like value of $\lambda$) result for perturbations, and this would be
presented in my talk. I didn't tell him what the result was actually (there was no blackboard nearby), and he didn't ask about it. He made no
comment to me at all. However, as you know, in two or three days, a new version of his paper had appeared with the correct dependence on $\lambda$
(up to a numerical coefficient for which he did not pretend). The latter coefficient was correctly and independently obtained in mine and Guth and
Pi papers (though in terms of different gauge-invariant quantities, the one which I used was $h=2{\cal R}$ in the present notation). Thus, one may have different guesses to what extent my firm statement had
affected Steven and had forced him to reconsider the first variant of his paper that he did very quickly, being a great scientist indeed.}''

This was a very stressful conference, where the new cosmological paradigm was developed. But we were young at that time. I remember a dinner at the house of Martin Rees, where I hypnotized Don Page and Dmitri Nanopoulos, but completely failed to hypnotize Hawking's secretary Judy Fella. And then I remember my surprise when Alesha suggested watching the famous movie ``Apocalypse Now'' by 	Francis Coppola. I asked why waste time watching movies during a conference like that?  In reply, Alesha gave me a brief lecture about it and convinced me to go. I knew that he was a great physicist, but I did not expect him to be a movie expert. We went to see it, and yes, he was right: the movie was dark, shocking, and powerful. 

Reflecting on that time, I recall how painful it was for me to hear from Alesha that my new inflation scenario, which I had been so proud of, had been ruled out. But without knowing it, I would not be bold enough to reconsider the basic principles of old and new inflation, and invent chaotic inflation \cite{Linde:1983gd}, which did not rely on the standard hot Big Bang theory. Thank you, Alesha!

And then he invented a stochastic approach to inflation \cite{Starobinsky:1986fx}, which was instrumental in the development of the theory of eternal chaotic inflation \cite{Linde:1986fc,Linde:1986fd,Goncharov:1987gl,Goncharov:1987ir,Linde:1993nz,Linde:1993xx}.  Then, together with Kofman and Starobinsky, we developed the theory of preheating, which is a crucial component of the theory of matter creation after inflation \cite{Kofman:1994rk,Kofman:1997yn,Greene:1997fu}.

But in this paper, I will mostly concentrate on the famous Starobinsky model \cite{Starobinsky:1980te}, its development, some of its generalizations, including $\alpha$-attractors, and the present status of these models in view of the recent observational data by ACT \cite{ACT:2025fju} and SPT~\cite{SPT-3G:2025bzu}.

\section{Starobinsky model and its different versions}\label{sec:star}

In most of the recent papers, the authors use the name ``Starobinsky model'' to describe, interchangeably,  one of the two different models,
\be\label{starR}
{{\cal L} \over \sqrt {-g} } = 
 {R\over 2} + {R^{2}\over 12 M^{2}}  \ ,
\ee
or 
\be\label{starf}
{{\cal L} \over \sqrt {- \tilde g} }=  {  \tilde R\over 2} - {1\over 2}\partial_{\mu}\vp \partial^{\mu} \vp -{3M^{2}\over 4} \left(1-e^{-\sqrt{2\over 3} \vp}\right)^{2} \ .
\ee
However, as we will discuss now, the relation between these two models is rather delicate.  Moreover, the original version of the Starobinsky model \cite{Starobinsky:1980te} was somewhat different.

 Instead of modifying the gravitational action as shown in \rf{starR}, Starobinsky in \cite{Starobinsky:1980te}  considered the Einstein equations with quantum corrections to the energy-momentum tensor,
\begin{eqnarray}\label{Star1980}
\langle T_{ik} \rangle &=&{1\over H_{0}^{2}}\Bigl(R_{i}^{l} R_{kl}- {2 \over  3} R R_{ik} - {1 \over 2} g_{ik} R_{lm}R^{lm} + {1 \over 4} g_{ik} R^{2}\Bigr) \\ \nonumber
& - & {1 \over 6 M^{{2} } }\, \Bigl(2R_{;i}^{;k} R_{kl} -    2g_{ik} R^{;l}_{;l}  -  2R R_{ik}  + {1 \over 2} g_{ik} R^{2}\Bigr) \ .
\end{eqnarray}
Here $H_{0}^{-2}= {k_{2} \over 2880\pi^{2}}$, $M^{{-2} }= -{k_{3} \over 2880\pi^{2}}$, where $k_{i}$ are the coefficients in the expression for the conformal anomaly. These numbers include the contributions of all particle species. A contribution of each type of particles is $O(1)$.
Consistency of the model required $M\ll 1$,  $H_{0} \ll 1$, $H_{0} \gtrsim 5 M$  \cite{Mukhanov:1981xt,Starobinsky:1983zz}.  During inflation, the Hubble constant decreased gradually, from $H_{0}$ to $M$, and then inflation ended.   

According to the original version of the Starobinsky model, the universe at $t = 0$ was in an unstable de Sitter state with the Hubble constant $H_{0}$. Since de Sitter universe is non-singular, one should be able to continue this solution to $t\to -\infty$. However, an unstable de Sitter state could not survive from $t = -\infty$ to $t = 0$, so inflation could not begin in an unstable state at $t = 0$.  Therefore, the original version of the Starobinsky model was incomplete.

Another problem was related to the amplitude of adiabatic perturbations in this model. To match the WMAP - Planck data, one would need to have $M \sim 1.3 \times 10^{-5}$, in Planck mass units $M_{p} = 1$ \cite{Mukhanov:1981xt,Starobinsky:1983zz, Kofman:1987ec}. But to achieve it, one would need to have more than $10^{{10}}$ different particle species contributing to the conformal anomaly.
Finding a realistic particle physics model with these properties is extremely difficult. Therefore in 1985, in our paper \cite{Kofman:1985aw}, Starobinsky abandoned the original idea of inflation due to radiative corrections and instead proposed using the action \rf{starR}.  

To explain the way from \rf{Star1980} to \rf{starR}, one should note that the first term in the energy-momentum tensor  \rf{Star1980}, proportional to  $H_{0}^{-2}$,  originates from the conformal anomaly, whereas the second term,  proportional to $M^{-2}$, could be either a result of a conformal anomaly, or of the term ${R^{2}\over 12 M^{2}}$ in the action   \rf{starR}, or both  \cite{Kofman:1987ec}.   Therefore, one can neglect all quantum corrections, thus removing the first term in \rf{Star1980}, and interpret the second term in \rf{Star1980} as a result of introducing the term ${R^{2}\over 12 M^{2}}$ in \rf{starR}.

I discussed these developments here because the standard reference to the Starobinsky model brings the readers to the original paper  \cite{Starobinsky:1980te}, which considered inflation as a result of quantum corrections to the Einstein theory, which required an extremely large number of particle species. This may lead to a misunderstanding, for example, when one attempts to evaluate the Starobinsky model from the perspective of swampland constraints on the number of species \cite{Lust:2023zql}.  That is why it is important to remember that in 1985 Starobinsky moved away from the idea of explaining inflation by quantum effects and started using the streamlined version of his model  \rf{starR} which did not rely on the large number of particle species.

From the point of view of the cosmological predictions, the transition from \rf{Star1980} to \rf{starR} was relatively uneventful, since inflation in the Starobinsky model \cite{Starobinsky:1980te} requires $H_{0}^{2}\gg  M^{2}$. Predictions of the original model \cite{Starobinsky:1980te} have a slight dependence on the value of $H_{0}$, but this dependence gradually disappears in the regime $H_{0} \gg 5 M$ \cite{Mukhanov:1981xt,Starobinsky:1983zz}. Under this condition,  the observational consequences of the original version of the Starobinsky model \cite{Starobinsky:1980te} almost coincide with the predictions of the model  \rf{starR}  \cite{Mukhanov:1981xt,Starobinsky:1983zz}. A detailed evaluation of these predictions, including both cases with and without conformal anomaly, can be found in the paper by Kofman, Mukhanov, and Pogosyan \cite{Kofman:1987ec}.

From a more general perspective, this modification of the Starobinsky model was substantial. In the original model \cite{Starobinsky:1980te}, inflation begins in a de Sitter state with the Hubble constant $H_{0}$, and $|R^{2}| = 12 H_{0}^{2} \ll 1$.  This solution disappears when one removes the first term in equation \rf{Star1980}. Thus, one could no longer argue that the universe was born in a dS state. However, inflationary solutions do exist  \cite{Barrow:1983rx, Kofman:1985aw}. The Hubble constant $H$ in the theory 
\rf{starR}  satisfies equation
\be
M^{2 }H^{2} = -6 H^{2} \dot H + \dot H^{2} -2H \ddot H \ .
\ee
During the slow-roll inflationary regime, one can drop the last two terms and find
\be
H = {M^{2} \over 6}  (t_{1}-t) \ .
\ee
Inflation ends at $t_{1}-t = O(1/M)$. By analogy with the chaotic inflation scenario, one could argue that inflation begins at $-t \sim M^{-2}$, i.e., at the Planck boundary when $H^{2} \sim |R| = O(1)$. This would provide a possible solution to the problem of initial conditions for this model. 
That is why Starobinsky compared his original model   \cite{Starobinsky:1980te} with new inflation, where inflation may begin in an unstable dS state \cite{Linde:1981mu}, and he compared its streamlined version \rf{starR} with chaotic inflation, which may begin at very large $\vp$,  close to the Planck density and curvature $R = O(1)$  \cite{Linde:1983gd}.

However, the situation with initial conditions in the model \rf{starR} is a bit tricky because of the term ${R^{2}\over 12 M^{2}}$. The gravitational action \rf{starR}  can be represented as ${R\over 2}(1 + {R\over 6M^{2}} )$. During inflation, the coefficient $(1 + {R\over 6M^{2}} )$ for a long time remains nearly constant, playing the role of an effective value of $M_{p}^{2}$, which is equal to $1$ at $R = 0$. But at large $R$, the effective value of  $M_{p}^{2}$ increases as ${R\over 6M^{2}}$.  For $R \gg M^{2}$, equation defining the Planck boundary  $R  \sim M_{p}^{2} \sim {R\over 6M^{2}}$ does not have any solutions.  This intuitive argument reveals a problem with the definition of the Planck boundary in the model \rf{starR}.  To analyze this issue properly, it is convenient to switch to the second formulation of the Starobinsky model given by \rf{starf}.

A transition from \rf{starR} to \rf{starf} took several years. First of all, after the metric transformation
${\tilde g_{\mu \nu} } = (1+{\phi\over 3M^{2}}) g_{\mu \nu}$ ,
the term $R^{2}$ disappears, and the model \rf{starR} transforms into a theory of a scalar field interacting with gravity   \cite{Whitt:1984pd}. Finally, a redefinition of the scalar field, making it canonically normalized \cite{Barrow:1988xh,Maeda:1987xf, Coule:1987wt}, transforms this theory into \rf{starf}. The transformation from \rf{starR} to \rf{starf} is often referred to as a transition from the Jordan frame to the Einstein frame.

In the Einstein frame, the definition of the Planck boundary is unambiguous: $\tilde R = O(1)$. Inflationary potential has a plateau with the height $V_{0} = {3\over 4} M^{2}$.  Inflation can only happen at the plateau with $V(\vp) \leq {3\over 4} M^{2}  \sim 1.3 \times 10^{{-10}}$.  Thus, inflation can never begin at the Planck boundary $\tilde R = 1$.

One may wonder whether inflation beginning at a density ten orders of magnitude below the Planck density is probable. To analyze it, one may calculate the probability of quantum creation of the Starobinsky universe ``from nothing". 

According to \cite{Linde:1983mx,Linde:1984ir,Vilenkin:1984wp}, the probability of quantum creation of a closed inflationary universe with potential $V(\phi)$ is given by
\be\label{prob}
P \sim  e^{-{24\pi^{2}\over V}} \ .
\ee
In application to the theory \rf{starf}, this yields  \cite{Vilenkin:1985md}
\be \label{probstar}
P \sim e^{-{32\pi^{2}\over M^{2}}} \sim 10^{-10^{12}} \ .
\ee
This makes the initial conditions for inflation in the Starobinsky model somewhat problematic.  

To alleviate this problem, one could try to add to the model  \rf{starR} a potential with a large energy density of the false vacuum. One could expect that this should increase $V$ in \rf{prob} and therefore increase the probability. However, Vilenkin has found that this does not increase the final result $P \sim e^{-{32\pi^{2}\over M^{2}}}$  \cite{Vilenkin:1985md}. 

One way to understand this surprising conclusion is to note that if we add a cosmological constant (false vacuum energy) to the Starobinsky model \rf{starR}, then, in the Einstein frame,  this would add to \rf{starf} not a constant but a term proportional to $e^{-2\sqrt{2\over 3} \vp}$, which is too steep to support inflation. This explains why adding false vacuum energy to the model \rf{starR} does not increase the probability of inflation  \rf{probstar}.

The simplest solution to this problem is to start not with the modified gravity model \rf{starR}, but directly with the model \rf{starf}, and add to it  the chaotic inflation potential ${m^{2}\over 2}\chi^{2}$:
\be\label{starchaot}
{{\cal L} \over \sqrt {- \tilde g} }=  {  \tilde R\over 2} - {1\over 2}\partial_{\mu}\vp \partial^{\mu} \vp -{3M^{2}\over 4} \left(1-e^{-\sqrt{2\over 3} \vp}\  \right)^{2} -{1\over 2}\partial_{\mu}\chi \partial^{\mu} \chi - {m^{2}\over 2} \chi^2\ .
\ee

\begin{figure}[H]
\centering
\includegraphics[scale=0.2]{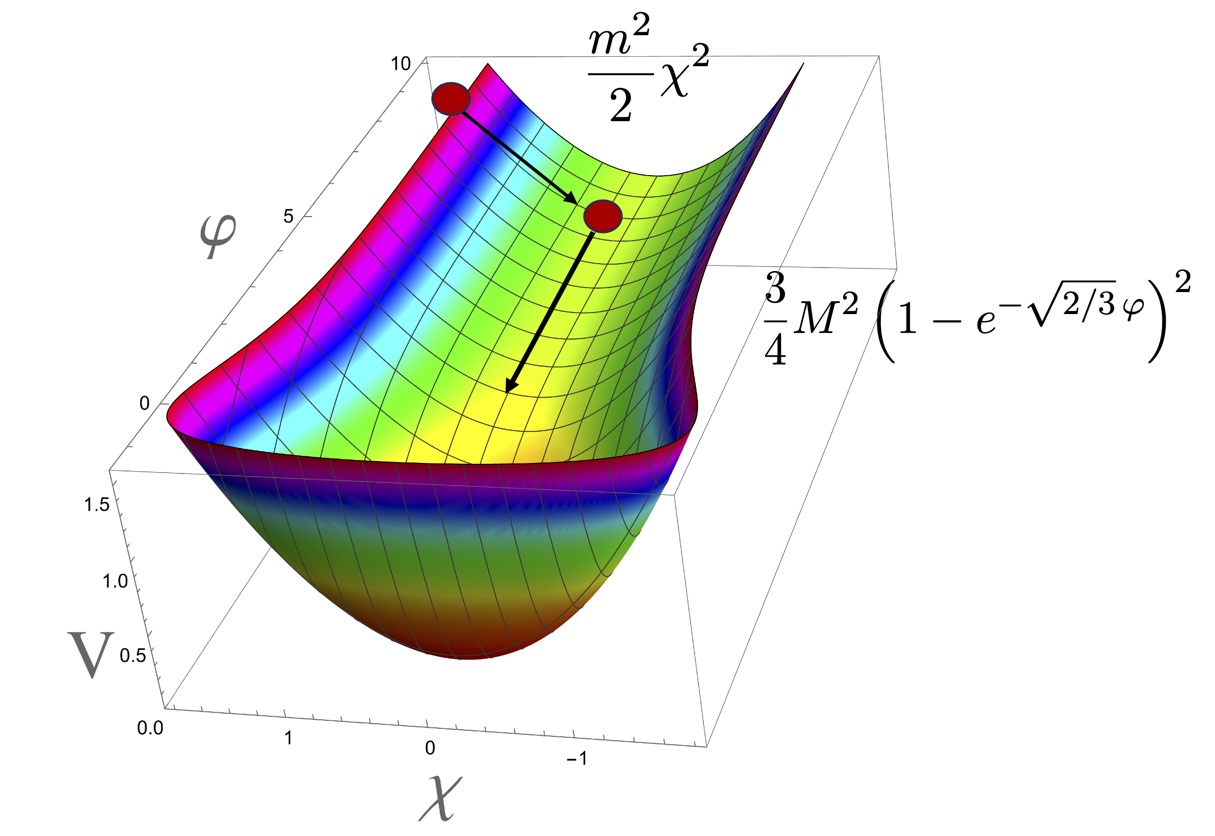}\vskip -5pt
\caption{\footnotesize  Inflation in the Starobinsky model \rf{starf} combined with the chaotic inflation mode  ${m^{2}\over 2}\chi^{2}$. The first stage of inflation begins at large $\chi$, as in the simple chaotic inflation model. After the field $\chi$ rolls to $\chi = 0$, the second stage of inflation begins, which is described by the model \rf{starf}. This solves the problem of initial conditions for inflation in the  model \rf{starchaot}.} 
\label{starinit}
\end{figure}

The potential of these two fields is shown in Fig. \ref{starinit}. The first stage of inflation in this scenario begins at $\chi \gg 1, \vp \gg 1$. At that stage, the potential is dominated by the chaotic inflation potential ${m^{2}\over 2}\chi^{2}$. This stage may begin at ${m^{2}\over 2} \chi^2 = O(1)$. According to  \rf{prob}, the probability of such a process is not exponentially suppressed \cite{Linde:1983mx,Linde:1984ir}.  During this first stage, the field $\chi$ moves down, but the field $\vp$ practically does not move.  After a long stage of inflation, the scalar field $\chi$ falls to $\chi=0$, and only then does inflation driven by the field $\vp$ begin, as shown in Fig.~\ref{starinit}. This second stage of inflation is responsible for the formation of the large-scale structure of the observable universe.  A similar scenario was described in \cite{Carrasco:2015rva,Dimopoulos:2016yep,Linde:2017pwt}  in application to $\alpha$-attractor models, and its validity was confirmed by a detailed numerical investigation in \cite{Corman:2022alv}.  

The Starobinsky model predicts \cite{Mukhanov:1981xt}
\be
\label{starpred}
 A_{s} = {V_{0} N_{e}^{2}\over 18 \pi^{2 }} \ , \qquad n_{s} = 1-{2\over N_{e}} \ , \qquad r = {12\over N^{2}_{e}} \ .
\ee 
where $V_{0} = {3M^{2}\over 4}$. 

The number of e-foldings $N_{e}$ depends on the process of reheating after inflation. To make the theory complete, including the process of reheating, it is necessary to add matter fields to it. One can do it, for example, by adding the Lagrangian of the standard model either to \rf{starR} or to \rf{starf}.   If one adds the standard model to \rf{starf}, it will be coupled to the field $\vp$ only via gravity. Meanwhile, if one adds the standard model to  \rf{starR}, the transformation \rf{trans} will modify the standard model and induce some terms describing its interaction with the inflaton $\vp$  \cite{Antoniadis:2025pfa}. 
The resulting two models, originating from \rf{starR} and \rf{starf}, and the process of reheating in these two models, differ from each other.

Another model, which provided a nearly perfect match to the Planck data, was the Higgs inflation model  \cite{Salopek:1988qh,Bezrukov:2007ep}.   The Higgs inflation model describes a scalar field with the standard Higgs potential, where the scalar field non-minimally is coupled to gravity, with the coupling $\sim \xi \phi^{2} R$ and $\xi \gg 1$. One can make a transformation to the Einstein frame where this term disappears, and the potential of the canonically normalized Higgs at its large values approaches a plateau exponentially fast. This potential is very similar to the potential in \rf{starf}, but it depends on $|\vp|$, so it has two plateaus, at $\vp \to \pm \infty$. For a discussion of initial conditions in this model, see \cite{Barvinsky:2022nfz}.

The Starobinsky model and the Higgs inflation model differ significantly from each other, yet both predict the same values for $n_{s}$ and $r$. There was some mystery surrounding it, so we sought to find a reason for this result and investigate whether other models may exhibit similar properties.

 \section{Conformal attractors}\label{tmod}

 The first class of a similar type that we were able to find had a rather unusual action \cite{Kallosh:2013hoa}
\begin{equation}
{\mathcal{L}\over \sqrt{-{g}}} =  {1\over 2}\partial_{\mu}\chi \partial^{\mu}\chi  +{ \chi^2\over 12}  R({g})- {1\over 2}\partial_{\mu} \phi\partial^{\mu} \phi   -{\phi^2\over 12}  R({g}) -{1\over 36} F\left({\phi/\chi}\right)(\phi^{2}-\chi^{2})^{2} \,.
\label{conform}
\end{equation}
where $F$ is an arbitrary function in terms of the variable $z = \phi/\chi$. 

This theory is locally conformal invariant under the  transformations  
\be \tilde g_{\mu\nu} = e^{-2\sigma(x)} g_{\mu\nu}\,
,\qquad \tilde \chi =  e^{\sigma(x)} \chi\, ,\qquad \tilde \phi =  e^{\sigma(x)}
\phi\ . \label{conf}\ee 

The field $\chi(x)$ is referred to as a conformal compensator.  Its kinetic term has a negative sign, but this is not a problem because it
can be removed from the theory by fixing the gauge symmetry
(\ref{conf}). In particular, one can use the gauge $\chi^2-\phi^2=6$ and resolve this constraint in terms of the fields 
$\chi=\sqrt 6 \cosh  {\varphi\over \sqrt 6}$,  $ \phi= \sqrt 6 \sinh {\varphi\over \sqrt 6} $ 
and the   field $\varphi$:~ ${\phi\over\chi} = \tanh{\varphi\over \sqrt 6}$. 
Our action \rf{conform} becomes
\begin{equation}\label{chaotmodel}
{\mathcal{L}\over \sqrt{-{g}}} =  \frac{1}{2}R - \frac{1}{2}\partial_\mu \varphi \partial^{\mu} \varphi -   F(\tanh{\varphi\over \sqrt 6})  \ .
\end{equation}
Note that at large $\vp$ the potential approaches a plateau
\be\label{plateau1}
V(\vp) = V_{0}(1- C  e^{-\sqrt{2\over 3\alpha} \varphi } )\ .
\ee
The constant $C$ can be absorbed into a redefinition of the field $\vp$. One can show that these models have the same predictions for $n_{s}$ and $r$ as the Starobinsky model for a very broad choice of the functions $F$. Moreover, for the special choice $F = {27 M^{2}\over (1+\chi/\phi)^{2}}$, the potential of this model exactly coincides with the potential of the Starobinsky model \rf{starf}.  

Models similar to  \rf{conform} are often used in supergravity. By using this approach, we developed one of the first consistent supergravity generalizations of the Starobinsky model  \cite{Kallosh:2013lkr,Kallosh:2013hoa}.

Thus, we have a large class of models, which have a different origin, not related to modified gravity \rf{starR}, but have very similar cosmological predictions. In this context, the model \rf{starf} is one of many models that equally well match the Planck data. Because of the conformal invariance of the models \rf{conform} and stability of predictions of these models with respect to large modifications of the function $F$, we called this class of models ``conformal attractors''  \cite{Kallosh:2013lkr,Kallosh:2013hoa,Kallosh:2013daa}.
The next step, the theory of $\alpha$-attractors, turned out to be even more interesting.
 
\section{\boldmath{$\alpha$-attractors}}\label{sattr}

The theory of $\alpha$-attractors \cite{Kallosh:2013hoa,Ferrara:2013rsa,Kallosh:2013yoa,Galante:2014ifa,Kallosh:2015zsa,Kallosh:2019eeu,Kallosh:2019hzo} has its roots in the hyperbolic geometry of the moduli space \cite{Kallosh:2015zsa}, and in supergravity \cite{Kallosh:2025jsb}. Here we will describe the phenomenological implications of two basic classes of single-field $\alpha$-attractors, T-models and E-models.   All of these models have supergravity generalizations.

\subsection{T-models}
The simplest T-model is given by 
 \be
{ {\cal L} \over \sqrt{-g}} =  {R\over 2}  -  {(\partial_{\mu} \phi)^2\over 2\bigl(1-{\phi^{2}\over 6\alpha}\bigr)^{2}} - V(\phi)   \,  .
\label{cosmoA}\ee
Here $\phi(x)$ is the inflaton.  In the limit $\alpha \to \infty$ the kinetic term becomes the canonical term $-  {(\partial_{\mu} \phi)^2\over 2}$.  One can recover the canonical normalization for any $\alpha$ by solving the equation ${\partial \phi\over 1-{\phi^{2}\over 6\alpha}} = \partial\vp$. The solution is $
\phi = \sqrt {6 \alpha}\, \tanh{\varphi\over\sqrt {6 \alpha}}$.
The full theory, in terms of the canonical variables  $\vp$,  becomes a theory with a plateau potential
 \be
{ {\cal L} \over \sqrt{-g}} =  {R\over 2}  -  {(\partial_{\mu}\varphi)^{2} \over 2}  - V\big(\sqrt {6 \alpha}\, \tanh{\varphi\over\sqrt {6 \alpha}}\big)   \,  .
\label{cosmoqq}\ee
 At large $\vp$, the potential can be represented as 
\be\label{plateau2}
V(\vp) = V_{0} - 2  \sqrt{6\alpha}\,V'_{0} \ e^{-\sqrt{2\over 3\alpha} \varphi } \ .
\ee
Here $V_0 = V(\phi)|_{\phi =  \sqrt {6 \alpha}}$ is the height of the plateau potential, and $V'_{0} = \partial_{\phi}V |_{\phi = \sqrt {6 \alpha}}$. The coefficient $2  \sqrt{6\alpha}\,V'_{0}$ in front of the exponent can be absorbed into a shift of the field $\varphi$. 

That is why all inflationary predictions in the regime with $e^{-\sqrt{2\over 3\alpha} \varphi } \ll 1$ are determined by two parameters, $V_{0}$ and $\alpha$,  and not by any other features of the potential $V(\phi)$:
\be
\label{pred}
 A_{s} = {V_{0}\, N_{e}^{2}\over 18 \pi^{2 }\alpha} \ , \qquad n_{s} = 1-{2\over N_{e}} \ , \qquad r = {12\alpha\over N^{2}_{e}} \ .
\ee 
Accuracy of these results increases with an increase of $N_{e}$ and a decrease of $\alpha$. These results are compatible with all presently available Planck/BICEP/Keck  data.

The amplitude of inflationary perturbations in these models matches the Planck normalization $A_{s} \approx 2.01 \times 10^{{-9}}$ for  $ {V_{0}\over  \alpha} \sim 10^{{-10}}$, $N_{e} = 60$, or for $ {V_{0}\over  \alpha} \sim 1.5 \times 10^{{-10}}$, $N_{e} = 50$.  For the simplest  model $V = {m^{2}\over 2} \phi^{2}$  one finds
\be\label{T}
V =  3m^{2 }\alpha \tanh^{2}{\varphi\over\sqrt {6 \alpha}} \ .
\ee
This simplest model is shown by the prominent vertical yellow band in Fig. 8 of the paper on inflation in the Planck2018 data release  \cite{Planck:2018jri}.  In this model,  the condition $ {V_{0}\over  \alpha} \sim 10^{{-10}}$ reads $ m  \sim   0.6 \times10^{{-5}}$. The small magnitude of this parameter accounts for the small amplitude of perturbations $A_{s} \approx 2  \times 10^{{-9}}$. No other parameters are required to match the presently existing Planck/BICEP/Keck data in this model. If the inflationary gravitational waves are discovered, their amplitude can be accounted for by the choice of the parameter $\alpha$ in \rf{pred}.
\subsection{E-models}
\begin{figure}[H]
\vskip -5pt
\centering
\includegraphics[scale=0.36]{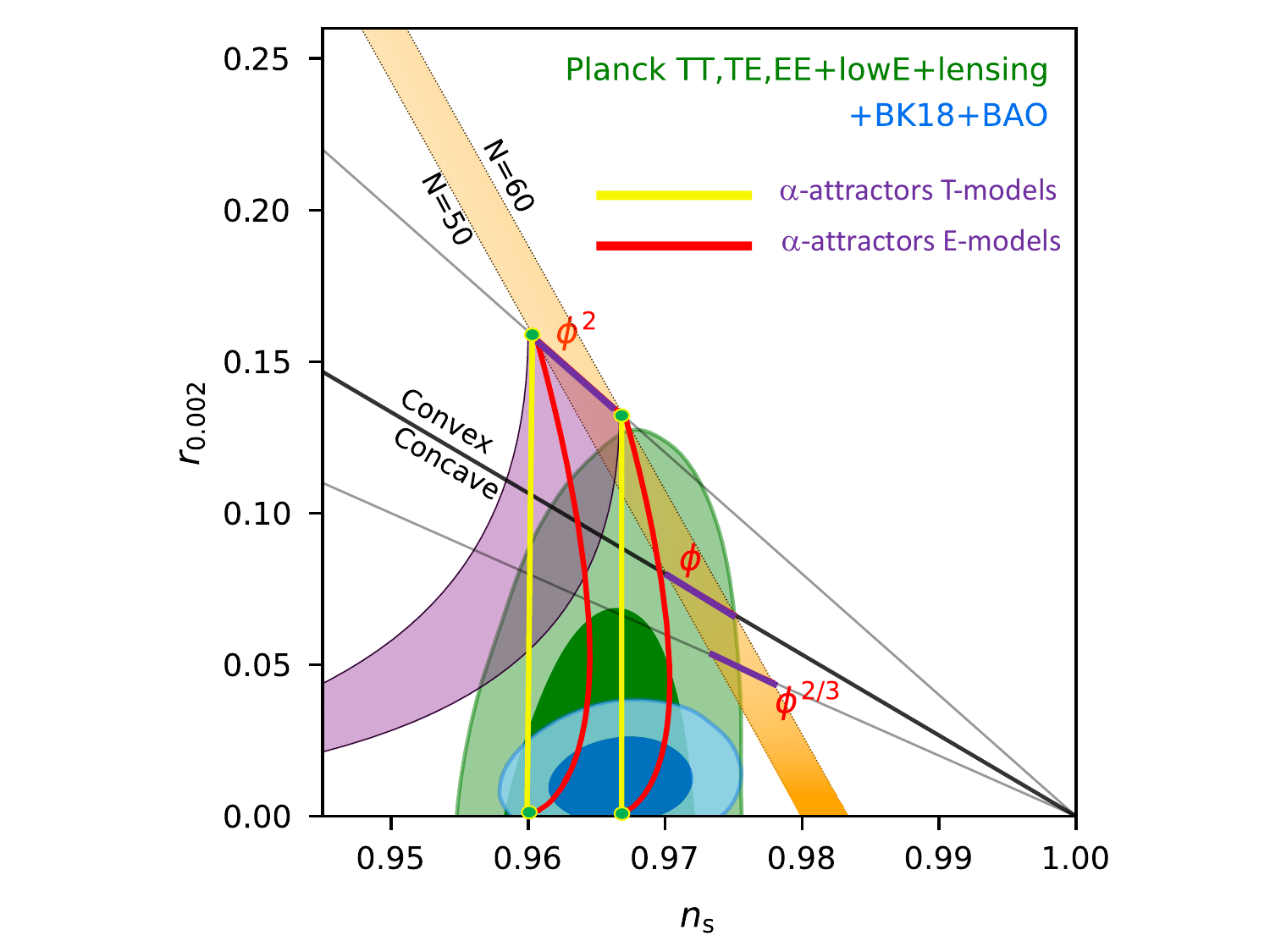}
\vskip -15pt
\caption{\footnotesize  BICEP/Keck results for $n_{s}$ and $r$ \cite{BICEP:2021xfz} superimposed with the predictions of the simplest $\alpha$-attractor T-model with the potential $\tanh^{2} {\varphi\over \sqrt{6\alpha}}$ and E-models (yellow lines for $N_{e} = 50, 60$) with the potential $\big (1-e^{-\sqrt{{2\over 3\alpha}}\varphi}\big )^{2}$ (red lines for $N_{e} = 50, 60$).}
\label{BICEPKeck0}
\end{figure}
The second family of $\alpha$-attractors,  called E-models ,is given by
 \be\label{actionE}
{ {\cal L} \over \sqrt{-g}} =  {R\over 2} - {3\alpha\over 4} \, {(\partial \rho)^2\over \rho^{2}}- V(\rho) \ .
\ee
As before, one can go to canonical variables,  which yields
\be\label{apole}
{ {\cal L} \over \sqrt{-g}} =  {R\over 2} - {1\over 2}  (\partial \varphi)^2- V(e^{-\sqrt {2\over 3\alpha}\varphi}). 
\ee
We consider $V(\rho)$  not singular at $\rho = 0$, which has a Minkowski minimum, e.g. $V(\rho) =V_{0}(1-\rho)^{2n}$. In  canonical variables, it becomes
\be\label{Emodel}
 V = V_0 \Bigl(1 - e^{-\sqrt {2 \over 3\alpha}\varphi} \Bigr)^{2n} . 
\ee
For $\alpha = 1$  and $n = 1$, this potential coincides with the potential of the Starobinsky model  \cite{Starobinsky:1980te}. In the small $\alpha$ limit, the predictions of the E-models coincide with the predictions of the T-models \rf{pred}, see Fig. \ref{BICEPKeck0}.

For a discussion of the problem of initial conditions in $\alpha$-attractor models, see e.g. \cite{Carrasco:2015rva,Linde:2017pwt,Corman:2022alv}.

\section{From Planck/BICEP/Keck to ACT, SPT and DESI}

In the CMB White Paper \cite{Chang:2022tzj} and LiteBIRD \cite{LiteBIRD:2022cnt} figures, one can see only 3 viable targets for B-mode searches: the Starobinsky model \cite{Starobinsky:1980te}, Higgs inflation \cite{Salopek:1988qh,Bezrukov:2007ep}, and  $\alpha$-attractors \cite{Kallosh:2013yoa}. 
\begin{figure}[H]
\centering
\includegraphics[scale=0.28]{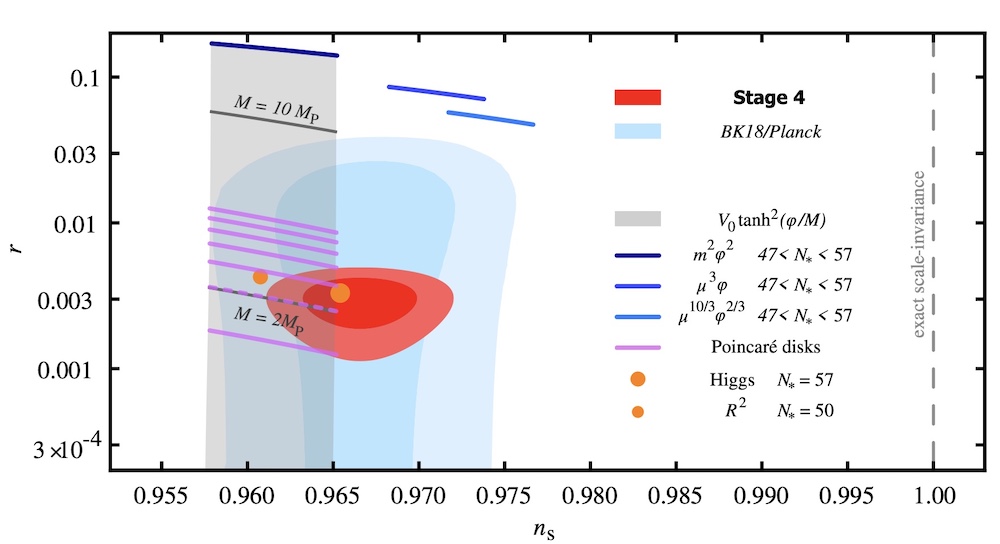}
\caption{\footnotesize The figure from  ``Snowmass2021 Cosmic Frontier: Cosmic Microwave Background Measurements White Paper'' \cite{Chang:2022tzj}. It shows the predictions of T-model $\alpha$-attractors with unconstrained values of $\alpha$ (gray area),  the predictions for $3\alpha= 7,6,5,4,3,2,1$ (purple lines), as well as Higgs inflation, $R^2$ inflation (red disks).  The predictions are for $47 < N_e< 57$.}
\label{Flauger0}
\end{figure}

With respect to the spectral index $n_{s}$, all of these models had a universal prediction $n_{s} = 1-2/N_{e}$. The typical value of $N_{e}$ depends on the mechanism of reheating after inflation and is expected to be in the range of 47 to 60. For example, for $N_{e} = 60$ one finds $n_{s} = 0.967$. 

The simplest reheating mechanisms in the Starobinsky model are not very efficient; therefore, one typically has $n_{s} \sim 0.962 - 0.963$. In the  Higgs inflation model, reheating is very efficient because the Higgs field is a part of the standard model, which is why one may expect $n_{s}\sim 0.966$. Reheating in $\alpha$-attractors is somewhat less constrained, depending on the specific model version. In particular, E-models introduced in  \cite{Kallosh:2013yoa} can reach $n_{s} \approx 0.97$ for $N_e = 60$, see Fig. \ref{BICEPKeck0}. There is also a class of quintessential $\alpha$-attractors, where the potential may not have a minimum. Such models may describe either a cosmological constant or dark energy, and they may have $n_{s}$ higher than that in more conventional models by $\Delta n_{s} \sim 0.006$ \cite{Dimopoulos:2017zvq,Dimopoulos:2017tud,Akrami:2017cir}.

The most recent constraint on $n_{s}$, based on the combination of all CMB-based data, including Planck. BICEP/Keck, ACT, and SPT, is $n_s = 0.9684\pm 0.003$, which is compatible with the inflationary models mentioned above. 

However, when the CMB data is combined with the results of DESI, the value of $n_{s}$ increases considerably. In particular, CMB data, including ACT data and DESI DR1 results, yield $n_s=0.9739\pm 0.0034$ \cite{ACT:2025fju,ACT:2025tim}. The latest SPT data, in combination with all other CMB results plus DESI DR2, give $n_s = 0.9728\pm 0.0027$ \cite{SPT-3G:2025bzu}, which is consistent with the ACT-DESI results.  As noted in \cite{ACT:2025tim}, these results disfavor the Starobinsky model, the Higgs inflation model, and many $\alpha$-attractors models at $\gtrsim 2\sigma$.

Thus, this result disfavored a large class of inflationary models, which had been favored by CMB data for more than a decade.  Indeed, according to \cite{ACT:2025fju,ACT:2025tim,SPT-3G:2025bzu}, the combination of CMB and DESI results eliminates all viable B-mode targets shown in figures by CMB-S4 and LiteBIRD collaborations.

However, the situation with the CMB-DESI related constraints is still highly uncertain. The authors of the ACT data release \cite{ACT:2025fju,ACT:2025tim}  in their talks emphasized that this is ``only $2\sigma$''. The discussion of these results in the SPT data release  \cite{SPT-3G:2025bzu} was even more cautious and nuanced. In particular, they noticed that the constraints  $n_s=0.9739\pm 0.0034$ \cite{ACT:2025fju,ACT:2025tim} and  $n_s = 0.9728\pm 0.0027$ \cite{SPT-3G:2025bzu} were obtained in the context of the $\Lambda$CDM theory, but in this context the  CMB data are in about $2.8 \sigma$ tension with the DESI DR2 data. 

The simplest way to understand it is to note that the CMB data alone are fully consistent with  $\Lambda$CDM, but the results of a combination of ACT and DESI DR2 are in more than $3\sigma$ tension with $\Lambda$CDM \cite{DESI:2025gwf}. Thus, one may wonder whether it is feasible to disfavor the Starobinsky model at $2\sigma$ level by combining data that are inconsistent with each other at the more than $3\sigma$ level in the $\Lambda$CDM context.  For a general discussion of combining mutually inconsistent data sets, see e.g. \cite{LuisBernal:2018drn}.

A more detailed discussion can be found in Section VII C ``Evaluating the consistency of CMB and BAO data in $\Lambda$CDM'' of the SPT data release  \cite{SPT-3G:2025bzu}. The authors advise caution in interpreting CMB+DESI data results in $\Lambda$CDM.

A subsequent discussion of this issue in \cite{Ferreira:2025lrd}  demonstrates that the fit of $\Lambda$CDM to CMB data exhibits a high degree of correlation
between $n_{s}$ and the BAO parameters $r_{d}h$ and $\Omega_{m}$. This correlation, and the tension in $r_{d}h$ and $\Omega_{m}$ between CMB data and DESI, can lead to significant shifts in $n_{s}$ when the two are combined. The authors concluded: ``Given the crucial role of $n_{s}$ in discriminating between inflationary models, we urge caution in interpreting CMB+BAO constraints on $n_{s}$ until the BAO-CMB tension is resolved.''
 
 Thus, for the time being, we may continue to use the Starobinsky model and Higgs inflation, and develop new, sophisticated versions of $\alpha$-attractors \cite{Kallosh:2024ymt,Carrasco:2025rud}. But it would be incorrect to simply discard the new data or wait until the dust settles. The first response to this situation was presented in \cite{Kallosh:2025rni}, where we developed the non-minimal chaotic inflation model \rf{nonminch}, which provides a good match to the ACT-DESI results. Shortly after that, approximately 50 papers on this issue were written, each offering its own approach to resolving the problem; see a recent review of this situation in \cite{Kallosh:2025ijd}. 

In particular, there is a large variety of Starobinsky-like models, which can be modified to match the ACT-DESI data (see, e.g., \cite{Antoniadis:2025pfa,Drees:2025ngb,Gialamas:2025ofz,Addazi:2025qra,Zharov:2025zjg,Ketov:2025cqg,Ellis:2025ieh}). Many of these proposals consider general versions of $f(R)$ gravity. A detailed discussion of such models and many related references can be found in 
\cite{Ketov:2025cqg}. A particular model described there contains additional terms, such as $R^{3}$ and $R^{4}$. $R^{6}$ with different coefficients in the range between $10^{{-5}}$ and $10^{{-9}}$.  Such generalizations of the Starobinsky model can match the CMB+DESI data at the expense of introducing additional extremely small parameters.
 
E-model $\alpha$ attractors with $\alpha \sim 6$ are only in $1 \sigma$ tension with the CMB-DESI related constraints for $N_{e} = 60$.  The predictions of quintessential  $\alpha$-attractors may also move closer to the CMB-DESI constraints. More considerable modifications can be achieved in the two-field models, such as the hybrid $\alpha$-attractors \cite{Kallosh:2022ggf}. In this class of models, one has two basic attractor regimes. If $V_{\rm up}$ is large and $\phi_{c} \ll  1$, then the last stage of inflation occurs as in the original hybrid inflation scenario,  with $n_{s}$  very close to 1. On the other hand, if uplifting is small, one has the standard $\alpha$-attractor result  $n_{s} = 1-{2\over N_{e}},  r = {12\over N^{2}_{e}}$.  Thus, by changing the parameters of the model, one may interpolate between $n_{s} = 1-{2\over N_{e}}$ and the  $n_{s} = 1$ \cite{Kallosh:2022ggf,Iacconi:2024hmg}.

The full theory of this effect is rather nuanced, as it depends on the relations between many parameters of the $\alpha$-attractor version of hybrid inflation models. For example, in certain cases, the process of spontaneous symmetry breaking can drive a short second stage of inflation, resulting in the production of large primordial black holes \cite{Braglia:2022phb}, which might even account for dark matter \cite{Garcia-Bellido:1996mdl}. 

Once one switches to the two-field models, other non-trivial models may emerge, such as the revised version of the curvaton scenario proposed in \cite{Byrnes:2025kit}.

\section{\boldmath Unity of cosmological attractors}\label{sec:pole}

To summarize the previous results: The Starobinsky model and the Higgs inflation model provide a very good match to the Planck data. There is a large class of the single-field models, such as conformal attractors and $\alpha$-arttractors,  which generalize the Starobinsky model and the Higgs inflation model. All of these models have predictions that are stable with respect to strong modifications of the potential.  In the previous section, we discussed a recent controversy that arises when combining CMB data and DESI results. If further investigations confirm the increase of $n_{s}$ found in \cite{ACT:2025fju,ACT:2025tim,SPT-3G:2025bzu}, then the stability of the predictions of the single-field models discussed above can make it difficult to modify them.  We discussed some of such modifications in the previous section, including switching to the multi-field models.

Going forward, we should also consider more substantial modifications of these models. All of these models have different motivations and different structures. One may wonder whether there is something common to all of them, ensuring stability of their predictions, and whether this common element may be preserved?

Indeed, there is a common element that is often overlooked in discussions of the Starobinsky model and Higgs inflation. We all know that one can begin with \rf{starR} and end up with \rf{starf}. However, we did not mention that halfway from one formulation to another,  one deals with scalar fields with singular kinetic terms, similar to the kinetic term $ {3\alpha\over 4} \, {(\partial \rho)^2\over \rho^{2}}$ in the $\alpha$-attractor action \rf{actionE}. 

Exploring the Higgs inflation model, one starts with terms like $\phi^{2} R$ and then switches to the Einstein frame. At the intermediate stages of calculations, one has a theory of a scalar field with a singular kinetic term of a similar type. Similar terms appear at the intermediate stage of calculations in conformal attractors. And finally, in $\alpha$-attractors these singular terms become a central part of the theory.

Note that in all of these models, the pole is of order 2. Such singular terms have an especially interesting interpretation for complex scalar fields, in terms of hyperbolic geometry,  famously illustrated in the artwork by Escher \cite{Kallosh:2015zsa}.

However, one may consider a more general situation, in the context of the so-called pole inflation models \cite{Galante:2014ifa}. It is obtained by generalizing the $\alpha$-attractor equation \rf{actionE}:
\be
{ {\cal L} \over \sqrt{-g}} =  {R\over 2} - {a_q\over 2} {(\partial \rho)^2 \over \rho^{q}} - V(\rho) \ .
\label{actionQ}\ee
Here, the pole of order $q$ is at $\rho=0$ and the residue at the pole is $a_q$. For $q = 2$, $a_2 = {3\alpha \over 2}$, this equation describes E-models of $\alpha$-attractors, but here we   consider general values of $q$. For $q \not = 2$ one can always rescale $\rho$ to make $a_q=1$.  Just as in the theory of $\alpha$-attractors, one can make a transformation to the canonical variables $\varphi$ and find that the asymptotic behavior of the potential $V(\varphi)$ during inflation is determined only by  $V(0)$ and the first derivative ${dV(\rho)\over d\rho}|_{\rho = 0}$. 

The potentials in this family of models also have inflationary plateaus at large $\vp$. For $\alpha$-attractors, the plateau of the potential is reached exponentially, but for $q > 2$ the approach to the plateau is controlled by negative powers of $\varphi$:
\be \label{BI1}
 V \sim V_{0}(1 -{m^{k}\over \vp^{k}}+... ) \ , \qquad n_{s} = 1-{2\over N_{e}}{k+1\over k+2} \ , \qquad {\rm where} \qquad k= {2\over 2-q} \ .
 \ee
Here are two simple examples of the potentials of this type and their cosmological predictions: 
\be\label{quart}
V = V_{0} \ {\varphi^{4}\over m^{4} + \varphi^{4}}\ , \qquad n_s =  1- {5 \over  3 N_{e}} \ , \qquad r = {4 m^{{4\over 3} }\over (3N_{e})^{5\over 3}} \ .
\ee
\be\label{D5}
V = V_{0} \ {\varphi^{2}\over m^{2} + \varphi^{2}}\ , \qquad n_s =  1- {3 \over  2 N_{e}} \ , \qquad   r=   {\sqrt 2\, m \over N_{e}^{3\over 2}} \ .
\ee
Some of these models described in  \cite{Kallosh:2018zsi} and \cite{Kallosh:2019eeu,Kallosh:2019hzo}   have an interpretation in terms of Dp-brane inflation \cite{Dvali:1998pa,Kachru:2003sx}. 
However, they also appear in other contexts. For example, the quadratic model $V  \sim   {\varphi^2\over m^2 + \varphi^2}$ was first proposed in \cite{Stewart:1994pt}. It was used in \cite{Dong:2010in} as an example of a flattening mechanism for the $\varphi^2$ potential due to the inflaton interactions with heavy scalar fields. 
To explain the basic idea, consider a potential
\be
V(\phi, \chi) =  \phi^{2}\chi^{2} + m^2 (\chi-\chi_{0})^{2} \ .
\ee
If $m^{2}$ is sufficiently large, during inflation driven by the field $\phi$, the field $\chi$ will track its instantaneous minimum, depending on $\phi$, and the effective potential of the field $\phi$ along the inflationary trajectory becomes
\be
V = m^{2} \chi_{0}^{2}\  { \phi^{2}\over   \phi^{2} +m^{2}}\  .
\ee
Potentials \rf{BI1} can also be obtained in the context of   polynomial $\alpha$-attractors   \cite{Kallosh:2022feu}, where
\be
{ {\cal L} \over \sqrt{-g}} =  {R\over 2} -{3\a\over 2} {(\partial \rho)^2 \over \rho^{2}} - V_0\,   {\ln^k\rho \over  \ln^k\rho+ 1 } \ .
\ee
By the change of variables  $\rho = e^{-\sqrt {2 \over 3\alpha}\varphi}$, as in  E-models \rf{apole}, \rf{Emodel},  this model is reduced to a model for a canonical scalar field with the potential 
\be
V =V_0 {|\varphi|^k\over \mu^k + |\varphi|^k} \ , \qquad \mu = \sqrt{3\alpha\,\over 2} \ .
\label{polynomial}\ee

Note that for any $k> 0$ the attractor values of $n_{s}$ in \rf{BI1} are greater than $1-2/N_{e}$, and in the limit $k \to 0$ one has $n_{s} \to  1-1/N_{e}$.
This means, for example, that for $N_{e} = 55$, this class of models can account for $n_{s}$ in the broad range $0.9636< n_{s} < 0.982$.

Yet another example is provided by the simplest chaotic inflation model with non-minimal coupling to gravity \cite{Kallosh:2025rni}:
 \begin{equation}
{1\over \sqrt{-g}} \mathcal{L} =  \tfrac12 (1+\phi) R  - \tfrac12 (\partial \phi)^2 -  \tfrac12  m^{2}\phi^{2}  \ . \label{JordanChaot} 
 \end{equation}
 As in the Starobinsky model and in Higgs inflation, one can switch to the Einstein frame with a canonically normalized scalar field minimally coupled to gravity. In the large $\vp$ limit one has
\be\label{nonminch}
V = {m^{2}\over 2} \left(1 -{8\vp^{-2}}+O(\phi^{-4})\right),
\ee
This model predicts, for $N_{e} = 60$,
\be\label{rchaot}
n_{s} = 0.9733, \qquad r = 0.0094 \ .
\ee
which matches the ACT and SPT constraints   $n_s=0.9739\pm 0.0034$ \cite{ACT:2025fju,ACT:2025tim} and  $n_s = 0.9728\pm 0.0027$ \cite{SPT-3G:2025bzu}. Note that  $r= 0.0094$ is almost 3 times higher than $r$ in the Starobinsky and Higgs inflation model, so it should be easier to test this model by future observations.

\begin{figure}
\centerline{\includegraphics[scale=.3]{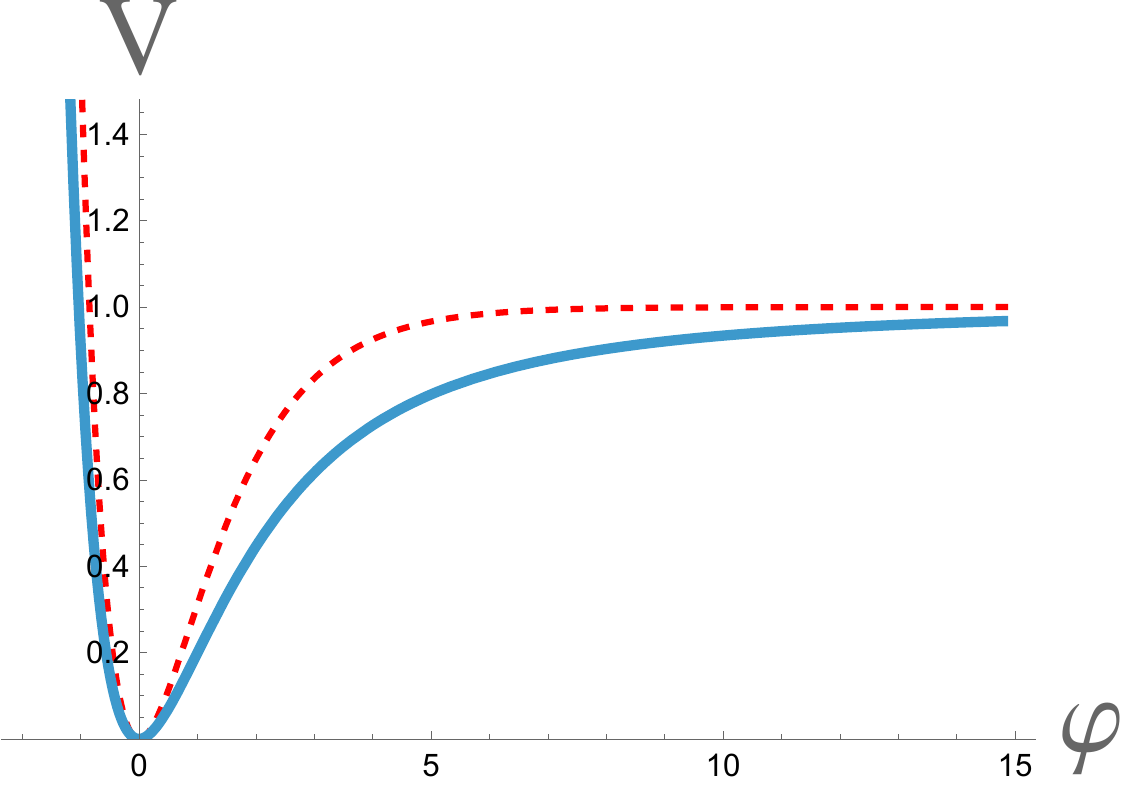}}
\caption{\footnotesize {The potential of the model \rf{JordanChaot} in the Einstein frame as a function of the canonical inflaton field $\vp$. As one can see, it is similar to the potential in the Starobinsky model  (red dashed line) and in the E-models of $\alpha$-attractors \cite{Kallosh:2013yoa}, but it approaches the plateau more slowly. }}
\vspace{-.3cm}
\label{pot}
\end{figure} 

An exact shape of the inflaton potential \rf{nonminch} of the non-minimal chaotic inflation model \rf{JordanChaot} is shown in Fig. \ref{pot}. It is interesting and perhaps somewhat unexpected that this potential looks remarkably similar to that of the Starobinsky model. The only difference is that the Starobinsky model potential approaches the plateau exponentially fast, whereas the potential of the non-minimal chaotic inflation approaches the plateau more slowly, following the inverse power law \rf{nonminch}. Both models are most economical, requiring only one parameter, but the Starobinsky model provides a good match to the Planck data, whereas the non-minimal chaotic inflation \rf{JordanChaot}  matches the CMB-DESI related constraints  \cite{ACT:2025fju,ACT:2025tim,SPT-3G:2025bzu}.

\section{Discussion}

In this paper, I have focused on a single topic, the Starobinsky model, because it is so popular, and this subject is particularly urgent in light of the recent CMB-DESI results. It is remarkable that a beautiful theoretical idea, proposed by Starobinsky 45 years ago, remains at the core of recent developments in both theoretical and observational cosmology. It is so unfortunate that we can no longer ask its author what he thinks about it now. I wish I could discuss his stochastic approach to inflation here, which was a stunning idea rediscovered and utilized by many. I remember how I was carrying in my backpack a draft of our paper with Lev Kofman and Alesha Starobinsky on reheating, first from Moscow to CERN, then to Stanford, until few yers later we had a chance to return to it, then completely re-write it, then throw this large paper to the garbage bin, and instead of it write a small paper on preheating. I still have on my computer a large draft of our paper with Kofman and Starobinsky on quantum cosmology, which has been left unchanged since 2009. There were many other urgent things to do, and we could certainly always call each other and return to it later. 

I wish...

\section*{Acknowledgement}
I am grateful to R. Bond, R. Kallosh, D. Roest, and E. Silverstein for many related discussions.   This work was supported by Leinweber Institute for Theoretical Physics at Stanford and by NSF Grant PHY-2310429.

\bibliographystyle{JHEP}
\bibliography{lindekalloshrefs}
\end{document}